\documentclass[cits,english]{PoS}

\usepackage{amssymb,fontenc,times,mathptmx,graphicx}
\usepackage{amsmath,latexsym}
\usepackage{axodraw,mathrsfs}
\usepackage{bbm}
\usepackage{macros}

\newcommand{\duzero}{\mcD\left[U_0\right]}

\title{The epsilon regime of chiral perturbation theory with Wilson-type fermions}

\ShortTitle{The epsilon regime of chiral perturbation theory with Wilson-type fermions}

\author{K. Jansen,\\
        NIC, DESY Zeuthen, Platanenallee 6 \\
        D-15738 Zeuthen, Germany\\
        E-mail: \email{karl.jansen@desy.de}}

\author{\speaker{A. Shindler}%
  \thanks{Current address:Instituto de F\'{\i}sica Te\'orica UAM/CSIC
Universidad Aut\'onoma de Madrid, Cantoblanco E-28049 Madrid, Spain}\\
        Theoretical Physics Division, \\
        Dept. of Mathematical Sciences, \\
        University of Liverpool \\
        Liverpool L69 7ZL, UK \\
        E-mail: \email{shindler@liv.ac.uk}}

\vspace*{-10mm}

\abstract{
  \begin{center}
    \includegraphics[draft=false,scale=1]{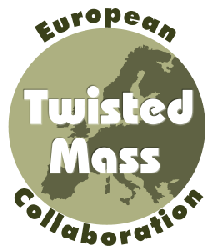}
  \end{center}
  \vskip 0.5cm
  In this proceeding contribution we report on the ongoing effort to simulate Wilson-type fermions 
  in the so called epsilon regime of chiral perturbation theory ($\chi$PT).
  We present results for the chiral condensate and the pseudoscalar decay constant 
  obtained with Wilson twisted mass fermions employing two lattice spacings, two different physical volumes and several quark 
  masses. With this set of simulations we make a first attempt to estimate the systematic uncertainties.
  \vskip 1.0cm
{\tt IFT-UAM/CSIC-09-53\\
DESY 09-186 \\
LTH 852}}

\FullConference{The XXVII International Symposium on Lattice Field Theory\\
                 July 26-31, 2009\\
                 Peking University, Beijing, China}

\begin{document}

\section{Introduction}
\label{sec:intro}
\vspace{-0.4cm}
\noindent The low energy dynamics of QCD can be quantitatively understood 
computing the universal low energy constants (LECs) of the chiral effective theory~\cite{Gasser:1983yg}.
QCD results obtained with simulations on a space-time lattice can be matched 
with the effective theory, if the range of quark masses and energy scales in the QCD computations,
are such that the higher order corrections in the effective theory calculations
are well under control.
A particular attractive framework in this respect is the so-called $\epsilon$-regime~\cite{Gasser:1987ah}.
In this regime the power counting for masses, momenta and linear size of the volume,
is such that higher order corrections are suppressed in comparison with the standard 
$p$-regime. Thus this regime provides an alternative, but also complementary,
way to determine the LECs in comparison with standard p-regime matchings.
The $\epsilon$-regime is properly matched if the volume in the QCD simulations, 
and correspondingly in the chiral perturbation theory ($\chi$PT) computations,
is larger than the confinement radius and if the Goldstone boson correlation
length of the system is smaller than the inverse linear size of the volume.
Typically the usage of Ginsparg-Wilson~\cite{Ginsparg:1981bj} (GW) fermions is preferred
because GW fermions have a natural definition for the topological charge, 
thus they allow to study the $\epsilon$-regime 
in a fixed topological sector~\cite{Hashimoto:2008fc} for which also
$\chi$PT formulae have been worked out~\cite{Damgaard:2002qe}. Wilson-type fermions can still, in principle, 
probe the $\epsilon$-regime, sampling all topological sectors~\cite{Jansen:2007rx,Jansen:2008ru}. An important point is the interplay between
low quark masses and a finite lattice spacing. It can be addressed
using the concept of generalized chiral expansions applied to the $\epsilon$-regime~\cite{Shindler:2009ri,Bar:2008th}.
A second issue is the algorithm which ought to be used to practically simulate in the $\epsilon$-regime
with Wilson-type fermions.
In these proceedings we cover these two topics, and present preliminary results
for the chiral condensate\footnote{In the text with $\Sigma$ we indicate, unless appearing
in renormalization group invariant combinations, the chiral condensate renormalized in the ${\overline{\rm MS}}$ scheme and a scale of 2 GeV.}
$\Sigma$ and the pseudoscalar decay constant $F$.
Additionally we discuss the systematic uncertainties which can affect
the determination of these LECs in the $\epsilon$-regime.
\vspace{-0.1cm}
\section{$\epsilon$ expansion with Wilson fermions}
\vspace{-0.4cm}
\label{sec:epsilon}
\noindent The $\epsilon$-regime has been introduced to cure the appearance of infrared
divergences in $\chi$PT when the mass squared of the Goldstone boson ($M_\pi$)
becomes smaller than the inverse size of the box ($V = L^3 \times T$).
The infrared divergences appear because the zero modes of the Goldstone bosons
are treated on the same footing as the non-zero modes. To cure this problem
Gasser and Leutwyler~\cite{Gasser:1987ah} proposed to change the power counting 
(from $p$-regime to $\epsilon$-regime) in this particular
region of the parameter space to achieve an exact resummation of the zero modes
and removing in this way any infrared divergence. As a result of this resummation
the order parameter of the chiral phase transition, the chiral condensate, vanishes
in the chiral limit. This signals the expected recovery of chiral symmetry when the quark mass
$m_{\rm q}$ vanishes {\em in a finite volume}, i.e. the absence of spontaneous chiral symmetry.
When studying the cutoff effects with Wilson-type fermions in the framework
of $\chi$PT the power counting of the $\epsilon$-regime in the continuum
\be
 \frac{1}{T} = {\rm O}(\epsilon) , \quad \frac{1}{L} = {\rm O}(\epsilon) , \quad M_\pi^2 = {\rm O}(\epsilon^4) \quad 
\left[ {\rm or} \quad m_{\rm q}={\rm O}(\epsilon^4) \right] ,
\ee
has to be augmented to include the lattice spacing $a$.
This is usually done connecting the power counting of the quark mass
$m_{\rm q}$ with the lattice spacing times the appropriate powers of the QCD scale $\Lambda$ to restore the 
proper dimensions.
Typically in the $p$-regime there are two different power countings depending on how the quark mass
and the lattice spacing are related: the GSM regime~\cite{Sharpe:2004ny} and the Aoki~\cite{Bar:2003mh}
or large cutoff effects (LCE) regime.
In the so-called GSM regime $m_{\rm q} \sim a\Lambda^2$ which implies $a = O(\epsilon^4)$.
In this regime the leading order (LO) cutoff effects can be reabsorbed
in the definition of the quark mass, implying no cutoff effects up to NLO order corrections
(this is indeed true also in the $p$-regime).
At NLO the chiral Lagrangian describing the dynamics of the Goldstone field $U(x)$ is given by
\bea
\mcL_{W\chi}^{(4)} &=& \mcL_{\chi}^{(4)} + a\widetilde{W} \Tr(\partial_\mu U^\dagger \partial_\mu U)\Tr(U + U^{\dagger})
- 2aB_0W \Tr(\mcM'^{\dagger} U + U^\dagger\mcM') \Tr( U + U^{\dagger}) + \nonumber \\
&-& a^2W' \big[\Tr( U + U^{\dagger})\big]^2 - 2aB_0H' \Tr( \mcM'+ \mcM'^\dagger ),
\eea
where $\mcL_{W\chi}^{(4)}$ is the continuum NLO chiral Lagrangian, $W$, $\widetilde{W}$ and $W'$ are LECs 
parametrizing cutoff effects, and $\mcM'$ is the shifted mass matrix which reabsorbes the LO O($a$) cutoff effects. 
Given our particular choice of the power counting it is easy to see that all
the corrections terms to the continuum Lagrangian are of NNLO~\cite{Shindler:2009ri,Bar:2008th}. This implies that Wilson fermions
are ``effectively'' free from discretization errors up to NNLO corrections.
This result is indipendent whether we use a clover term in the fermion action or not and whether
we use twisted mass fermions or not. 

The Aoki regime is defined by $m_{\rm q} \sim a^2\Lambda^3$ which implies $a = O(\epsilon^2)$.
The LO chiral Lagrangian in the Aoki regime is 
\bee
\mathcal{L}_{W\chi}^{(2)} = \frac{F^2}{4}
\Tr\left[ \partial_\mu U^\dagger \partial_\mu U \right] - \frac{\Sigma}{2}
\Tr\left[ \mcM'^\dagger U + \mcM' U^\dagger \right] - a^2W'
\left[ \Tr \left( U + U^\dagger\right)\right]^2.
\label{eq:Wchi2}
\eee
It contains already at LO O($a^2$) cutoff effects that cannot be reabsorbed
in the definition of the quark mass. 
It is well known~\cite{Aoki:1984qi,Sharpe:1998xm} that these LO cutoff effect in infinite volume 
change the vacuum structure of the effective theory leading to two 
possible scenarios for the chiral phase diagram~\cite{Farchioni:2004us}.
In the Aoki scenario the pattern of spontaneous symmetry breaking 
changes the continuum $SU(2)_L \times SU(2)_R \rightarrow SU(2)_V$
into $SU(2)_V \rightarrow U(1)$ signalling the spontaneous breaking of flavour (and parity) 
symmetry.
In the Sharpe-Singleton scenario there is no phase transition in the chiral limit
and the Goldstone bosons remain massive.
It is quite clear that in both scenarios the physics of the zero-modes is quite different
from the one of the continuum or of the GSM regime.
More work is needed in order to understand Wilson-type fermions in the deep
chiral regime.

The $\epsilon$-regime gives us the possibility to have a transition region between
the two regime in which the cutoff effects appear at NLO~\cite{Shindler:2009ri,Bar:2008th}.
To understand this regime from a power counting point of view we can decide
to set $a=O(\epsilon^3)$. 
The LO action contains, as in the continuum, the mass term $\left[S_2^{(0)}\right]_{\mcM}$ and the kinetic term for the non-zero-modes.
First corrections due to a finite lattice spacing appear at NLO. The partition function
of the effective theory at NLO can be written as
\be
\mcZ = \mcN \int \duzero {\rm e}^{-\left[S_2^{(0)}\right]_{\mcM}\left(\Sigma_{\rm eff}\right)} \times Z_\pi\left[U_0\right], \qquad \left[S_2^{(0)}\right]_{\mcM} = -\frac{\Sigma}{2}\int d^4x \Tr \left[\mcM'^{\dagger} U_0+ U_0^{\dagger} \mcM'\right], 
\label{eq:znlo}
\ee
where\footnote{$\Sigma_{\rm eff}$ contains the NLO correction to $\Sigma$ coming from the one loop non-zero modes contribution.}
\be
Z_\pi\left[U_0\right] = \mcN\left\{1 + W'a^2V\left[\Tr\left(U_0+U_0^\dagger\right)\right]^2 +
\frac{2 a W\Sigma V}{F^2}\Tr\left[\mcM'^\dagger U_0 + U_0^\dagger\mcM'\right]\Tr\left[U_0 + U_0^\dagger\right] \right\}.
\label{eq:zpinlo}
\ee
It is clear from eqs.~\eqref{eq:znlo} \eqref{eq:zpinlo} that the discretization errors of
O($a^2$) are of NLO while the discretization errors of O($am_{\rm q}$) are of higher order being of O($\epsilon^3$).
With this partition function it is straightforward to introduce appropriate sources
and compute two-point functions.
We refer to~\cite{Shindler:2009ri,Bar:2008th} for details on the computation.
It turns out that the relative O($a^2$) cutoff effects corrections, over a wide range of values 
for $m_{\rm q}\Sigma V$, are at most of few percent.
Another result of our analysis~\cite{Shindler:2009ri,Bar:2008th} is that there are appropriate linear combinations of correlation
functions like $C_{\rm S}(x_0)/4+3C_{\rm P}(x_0)$ or $C_{\rm AA}(x_0) + C_{\rm VV}(x_0)$
which are free from O($a^2$) effects and have leading cutoff effects of O($am_{\rm q}$).
\enlargethispage*{2.01\baselineskip}
We can summarize the results of our analysis generalizing the $\epsilon$ expansion using
Wilson fermions in the following way.
In the Aoki regime there could be large cutoff effects and more work is needed
to completely understand the interplay between quark mass and lattice spacing effects.
In the GSM regime Wilson fermions have no cutoff effects up to NNLO.
In the transition region between the two regimes we have a tool to analyze cutoff effects
and we have computed correlators including O($a^2$) and O($am_{\rm q}$).
The proper power counting has been identified as $a =O(\epsilon^3)$
and two-point functions have been computed up to relative O($\epsilon^3$) corrections.
We have noticed that certain linear combinations have no discretization
errors up to NNLO.
This analytical effort obviously has to be combined with numerical simulations 
which can tell us which regime has been properly matched.
We are currently extending this computation for Wilson-twisted mass fermions~\cite{Bar:2009ip}.
\vspace{-0.1cm}
\section{Numerical results with Wilson twisted mass fermions}
\label{sec:numerics}
\begin{figure}
\vspace{-0.7cm}
\begin{minipage}[t]{0.5\linewidth}
\centering
  \includegraphics[width=0.75\linewidth,angle=270]{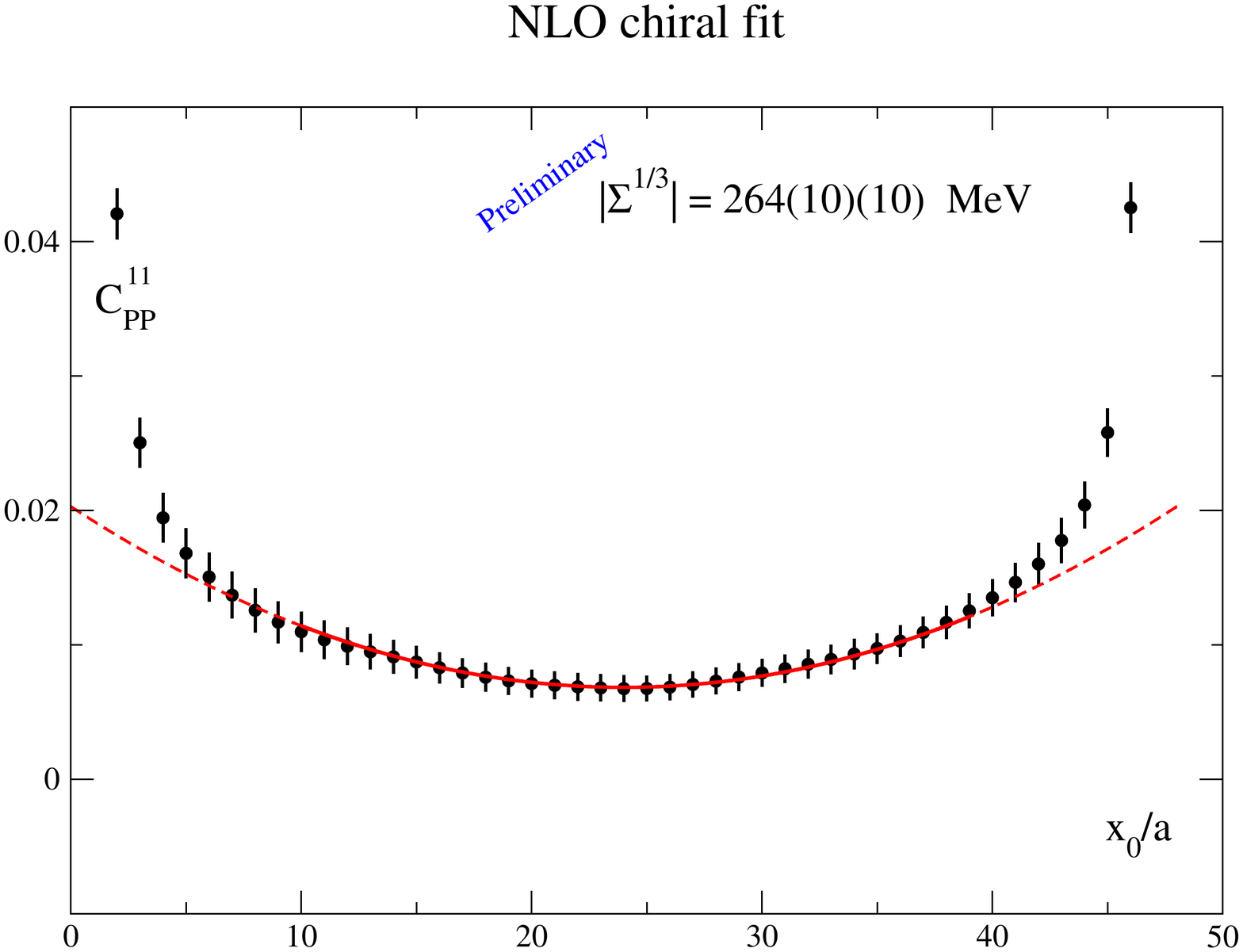}
\caption{Euclidean time dependence of the pseudoscalar density two-point function
and the fit results obtained using the NLO formula
in the continuum. The fit curve becomes dashed outside the fit range.}
\label{fig:pp24mu39}
\end{minipage}
\hspace{0.5cm}
\begin{minipage}[t]{0.5\linewidth}
  \centering
  \includegraphics[width=0.8\linewidth,angle=270]{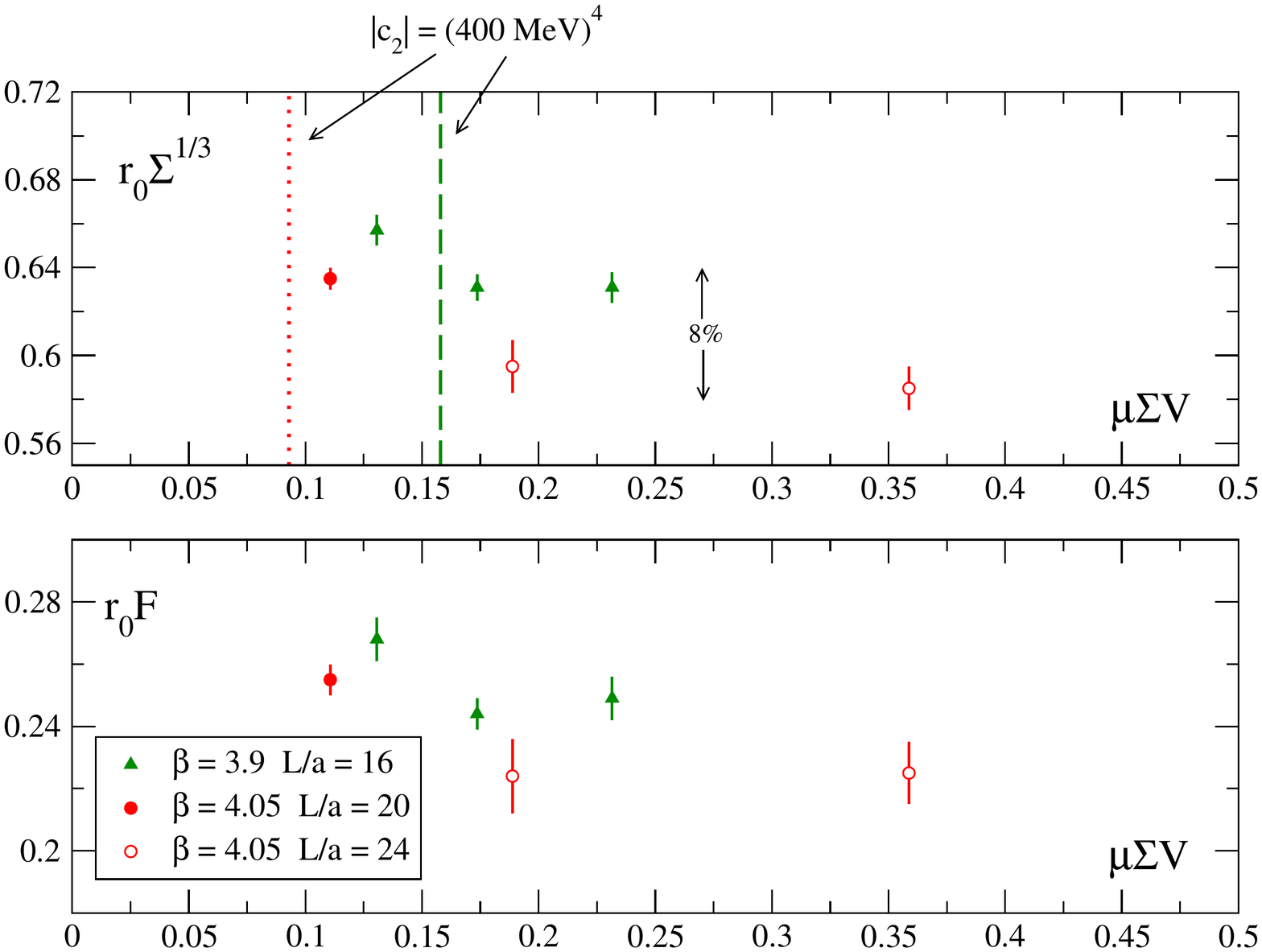}
\vspace{-0.5cm}
  \caption{Plot of $r_0|\Sigma|^{1/3}$ and $r_0 F$ for all the simulations point we have at the moment as a function
of $\muq \Sigma V$.}
  \label{fig:summary}
\end{minipage}
\end{figure}

\vspace{-0.4cm}
\noindent To perform simulations in the $\epsilon$-regime with Wilson-type fermions
we need specific algorithmic improvements.
The main ingredient for these improvements is {\it reweighting}.
We first proposed to use the PHMC algorithm~\cite{Frezzotti:1997ym} combined with exact reweighiting
to include in an exact way the low modes of the Wilson twisted mass operator~\cite{Jansen:2007rx}.
Different but somehow related techinques have been later proposed based
on stochastic reweighting in the standard mass~\cite{Hasenfratz:2008fg} or in the twisted mass~\cite{Luscher:2008tw}.
The main goal of reweighting in the $\epsilon$-regime is to ensure a better
sampling of the configuration space and to avoid instabilities issues
with HMC-like algorithms. The usage of twisted mass is particularly beneficial because
it provides a sharp infrared cutoff for the spectrum of the lattice operator.
We have performed simulations with a tree-level improved gauge action~\cite{Weisz:1982zw}
and $N_{\rm f} = 2$ Wilson twisted mass fermions~\cite{Frezzotti:2000nk,Frezzotti:2003ni,Shindler:2007vp}.
We summarize in tab.~\ref{tab:sim_par} the parameters of the numerical simulations
we are currently performing. To determine the LECs we compute two-point functions and we 
compare the Euclidean time dependence of the correlation functions with the 
time dependence predicted by $\chi$PT.
In fig.~\ref{fig:pp24mu39} we show the numerical results, for the ensemble $F_2$, for the 
charged pseudoscalar density two-point function and the fit results obtained using the NLO formula
in the continuum~\cite{Hasenfratz:1989pk,Hansen:1990un}
\bee
C_{\rm P}(x_0) = \frac{\Sigma_{\rm eff}^2}{3}\left\{\frac{X_2(z_{\rm eff})}{X_1(z_{\rm eff})} + \frac{3}{F^2}\left( 1 - \frac{1}{3}\frac{X_2(z)}{X_1(z)}\right)
\frac{T}{L^3}h_1(x_0/T) \right\} ,
\qquad z_{\rm eff} = 2 \mu_{\rm q} \Sigma V \left(1-\frac{N_f^2-1}{N_fF^2}\bar{G}(0)\right) , 
\label{eq:PP}
\eee
(see ref.~\cite{Shindler:2009ri} for unexplained notations) in the fit range $10 < x_0/a < 38$.
It turns out that the fit results are very stable if we change the number of data points included in the fit.
We have repeated this analysis for all the simulation points of tab.~\ref{tab:sim_par}
and the results for the chiral condensate $\Sigma$
and the pseudoscalar decay constant $F$ have been collected in fig.~\ref{fig:summary}.
\begin{table}[t]
\begin{minipage}[b]{0.5\linewidth}
\vspace{-2.5cm}
 \begin{tabular}{|cccccc|} \hline
lattice  & $\beta$ & Lattice   & $a$ [fm] & $a\muq$  & $N_{\rm traj}$    \\ \hline
${\rm I}_1$ & $3.9$ & $16^3 \times 32$   & $0.079$ & $0.0005$ &  $2500$ \\
${\rm I}_2$ & $3.9$ & $16^3 \times 32$   & $0.079$ & $0.00075$ & $3500$ \\
${\rm I}_3$ & $3.9$ & $16^3 \times 32$   & $0.079$ & $0.001$ &  $3135$ \\
${\rm F}_1$ & $4.05$ & $24^3 \times 48$  & $0.063$ & $0.00039$ &  $1755$ \\
${\rm F}_2$ & $4.05$ & $24^3 \times 48$  & $0.063$ & $0.00078$ &  $2316$ \\
${\rm F}_3$ & $4.05$ & $20^3 \times 40$  & $0.063$ & $0.00039$ &   $2500$ \\
 \hline
 \end{tabular}
\caption{Simulation parameters of the runs performed where $\beta=6/g_0^2$, the lattice spacing $a$~\cite{Dimopoulos:2009ip}, 
$\muq$ is the twisted mass parameter and $N_{\rm traj}$ is the number
of trajectories (with unity trajectory length) excluding the thermalization process.} 
\label{tab:sim_par}
\end{minipage}
\hspace{1.0cm}
\begin{minipage}[b]{0.5\linewidth}
  \begin{tabular}{|c|c|c|} \hline
    Group  &
    $r_0\Sigma^{1/3}$ & $r_0F$   \\ \hline
    This work   & $0.595(12) $ & $ 0.224(12)$ \\
    ETMC~\cite{Dimopoulos:2009ip}   & $0.574(28)$ & $ 0.183(8) $  \\
    HHS~\cite{Hasenfratz:2008ce}  & $0.617(15)$ & $0.224(10)$\\
    JLQCD~\cite{Fukaya:2007pn}   & $0.596(10)$ & $ 0.217(14)$  \\
    JLQCD~\cite{Fukaya:2007yv}   & $0.624(17)(27)$ & ----- \\
    \hline
  \end{tabular}
\caption{Table comparing the results presented in this proceedings with
recent results obtained in the $\epsilon$-regime
using $N_{\rm f} = 2$ overlap fermions and clover-type fermions. The result obtained in the $p$-regime 
with $N_{\rm f} = 2$ Wilson twisted mass fermions by ETMC
is also added for comparison.}
\label{tab:results}
\end{minipage}
\end{table}
This figure summarizes the results of simulations at 2 values of the lattice spacings,
2 different physical volumes and several values of the quark masses, thus allowing us to attempt
a first understanding of the systematic errors.
Together with the numerical data we also plot two vertical dashed ($\beta = 3.9$) and dotted
($\beta=4.05$) lines, 
for the two different lattice spacings, indicating the value
of $\mu_{\rm q}\Sigma V$ where we enter the so called Aoki regime (see sect.~\ref{sec:epsilon}).
This value depends on the value of an unknown low energy constants, called $c_2 \propto -W'$ 
which parametrizes the O($a^2$) effects. The two lines are the bound for $\mu_{\rm q}\Sigma V$
for two different lattice spacings given the indicative, but plausible, value $|c_2| = (400 {\rm MeV})^4$.
The figure can be interpreted in the following way. The two different lattice spacings
and same physical volumes (filled simbols) agree with themselves if we exclude the
most chiral point at $\beta = 3.9$ which is beyond the dashed line we have drawn.
This might indicate that cutoff effects might be small if we keep the value of $\mu_{\rm q}\Sigma V$
larger than the O($a^2 W' V$). The second effect is visible when we change the physical volume.
There seems to be a discrepancy between the two volume simulated both for 
the chiral condensate and the decay constant.
This discrepancy might indicate the inadequacy of a NLO fit when the volume is not large enough.
On the other side there is a good consistency in the LECs if we change the value of the mass 
at the largest volume available for both lattice spacings (excluding the most chiral point already dicussed).
If we take the finest lattice spacing and the largest volume as our best estimate of the LECs
we obtain the preliminary results 
\bee
r_0\Sigma^{1/3} = 0.595(12), \qquad r_0F = 0.224(12) 
\eee
In tab.~\ref{tab:results} we compare our determination for the LECs with other determinations
obtained in the $\epsilon$-regime~\cite{Hasenfratz:2008ce,Fukaya:2007pn,Fukaya:2007yv} and with the determination obtained by ETMC in the 
$p$-regime. We observe a very good agreement with all the determinations for the chiral condensate.
For the pseudoscalar decay constant we observe a good agreement among all the determinations
in the $\epsilon$-regime, while there is some tension with the ETMC determination in the 
$p$-regime~\cite{Dimopoulos:2009ip}.
\vspace{-0.1cm}
\section{Conclusions}
\label{sec:conclu}
\vspace{-0.4cm}
\noindent Probing the $\epsilon$-regime with Wilson-type fermions is possible.
We have now an analytical tool which allows us to study the combined
lattice spacing, volume and mass dependence of correlation functions in the $\epsilon$-regime,
and we have computed two-point functions at NLO.
We have performed several simulations in the $\epsilon$-regime with Wilson twisted mass,
using a PHMC algorithm combined with exact reweighting. This allows
in principle the determination of the LECs $\Sigma$ and $F$ without contaminations
from chiral logs.
Systematic uncertainties can now be addressed because of the rather large set
of simulations we have performed, indicating an $8\%$ of systematic errors 
coming from a too small volume and a $4-10\%$ of systematic errors coming from 
the finite cutoff.
To ensure a reliable understanding of the systematic uncertainties and thus remove them
we will enlarge the set of simulation points.
We will also try to enlarge the set of physical quantities to compute
to fully take advantage of the expensive dynamical simulations perfomed so far.
\enlargethispage*{2.01\baselineskip}
From the analytical side we are curently extending the NLO computation performed
with Wilson fermions to Wilson twisted mass~\cite{Bar:2009ip}.
We see no reason why this computation could not be done with staggered fermions, and moreover it could be 
a good corner of the parameter space where to test the rooting approach.
\vspace{-0.4cm}
\acknowledgments
\vspace{-0.4cm}
\noindent We thank the organizers of ``Lattice 2009'' for the very interesting
conference realized in Beijing. We also aknoweledge computer time 
made available by CNRS on the BlueGene system at IDRIS and CCIN2P3
in Lyon, by the NW-Grid in UK and by University of Liverpool.
A.S. thanks the Spanish Consolider-Ingenio 2010 
Programme CPAN (CSD 2007-00042) and Comunidad Aut\'onoma de Madrid, CAM under grant HEPHACOS P-ESP-00346
for funding.

\bibliographystyle{JHEP-2}    % if you use h-elsevier.bst
\bibliography{latt_09}      % or whatever your .bbl file is

\end{document}